\documentclass[a4paper]{jpconf}

\usepackage{amssymb}

\usepackage{color}    

\newcommand{\al}{\alpha}
\newcommand{\pa}{\partial}

\newcommand{\om}{\omega}

\newcommand{\De}{\Delta}

\newcommand{\rar}{\rightarrow}

\newcommand{\non}{\nonumber}
\begin{document}

\title[Two-body Coulomb problem and hidden $g^{(2)}$ algebra: cubic polynomial algebra]{Two-body Coulomb problem and hidden $g^{(2)}$ algebra: superintegrability and cubic polynomial algebra}

\author{Alexander~V.~Turbiner$^1$ and Adrian M. Escobar-Ruiz$^2$}
\address{$^1$Instituto de Ciencias Nucleares, Universidad Nacional
Aut\'onoma de M\'exico, Apartado Postal 70-543, 04510 M\'exico,
D.F., Mexico}
\address{$^2$Departamento de F\'isica, Universidad Aut\'onoma Metropolitana-Iztapalapa,
Apartado Postal 55-534, 09340 M\'exico, D.F., Mexico}
\ead{turbiner@nucleares.unam.mx, admau@xanum.uam.mx}

\vspace{10pt}
\begin{indented}
\item[]October 2023
\end{indented}

\begin{abstract}
It is shown that the two-body Coulomb problem in the Sturm representation leads to a new two-dimensional, exactly-solvable, superintegrable quantum system in curved space with a $g^{(2)}$ hidden algebra and a cubic polynomial algebra of integrals. The two integrals are
of orders two and four, they are made from two components of the angular momentum and from the modified Laplace-Runge-Lenz vector, respectively. It is demonstrated that the cubic polynomial algebra is an infinite-dimensional subalgebra of the universal enveloping algebra $U_{g^{(2)}}$.
\end{abstract}

%
\vspace{2pc}

\vspace{2pc}

\section{Introduction}

In the talk given at QTS12 by the second author (A.M.E.R.) and based partly on \cite{TE:2023}
the two-body Coulomb problem in the three-dimensional space
of relative motion is considered, where the Hamiltonian is of the form
\begin{eqnarray}
\label{Hamiltonian}
\hat {\cal H} \ = \ \frac{1}{2\,\mu}\,{\hat{\bf  p}^2}\ - \ \frac{\alpha}{r} \ ,
\end{eqnarray}
here $\al > 0$ for the case of charges of opposite signs, $\hat{\bf  p} = -i\, \hbar\, \nabla$ is the momentum operator, $r=\sqrt{x^2+y^2+z^2}$; for simplicity a unit reduced mass, $\mu=1$ and $\hbar=1$ is assumed from now on.

\noindent It is well-known that the angular momentum
\begin{eqnarray}
\label{L}
\hat{\bf  L} \ = \ {\bf r} \times \hat{\bf  p} \ ,
\end{eqnarray}
and the Laplace-Runge-Lenz vector
\begin{eqnarray}
\label{A}
\hat{\bf  A} \ = \ \frac{1}{2}(\,\hat{\bf  p} \times \hat{\bf  L}\,-\,\hat{\bf  L} \times \hat{\bf  p}\,) \ - \ \frac{\alpha}{r}\,{\bf r} \ ,
\end{eqnarray}
are integrals of motion: they commute with the Hamiltonian (\ref{Hamiltonian}),
\[
[\hat{\bf  L} ,\,\hat {\cal H}]\ = \ [\hat{\bf  A} ,\,\hat {\cal H}]\ = \ 0\ .
\]
Hence, the system (1) is maximally superintegrable.
It is easy to verify the existence of the well-known relation
\begin{eqnarray}
\label{A2}
\hat{\bf  A}^2 \ = \ \al^2 \ + \ 2 \,\hat {\cal H}\,(\,\hat{\bf L}^2 \ + \ 1\,)\ ,
\end{eqnarray}
and constraint
\begin{equation}
     \hat{\bf  L}\,\cdot\,\hat{\bf A} \ = \ \hat{\bf  A}\,\cdot\,\hat{\bf L} \ = \ 0 \ .
\end{equation}
The integrals $\hat{\bf  L},\,\hat{\bf  A}$ obey the following commutation relations:
\begin{eqnarray}
\label{LA-comm-Rel}
[\hat{L}_i \,,\,\hat {L}_j] \ = \ i\,\epsilon_{ijk}\,\hat{L}_k \quad , \quad [\hat{A}_i \,,\,\hat {L}_j] \ = \ i\,\epsilon_{ijk}\,\hat{A}_k \quad , \quad  [\hat{A}_i \,,\,\hat {A}_j] \ = \ -2\,i\,\epsilon_{ijk}\,\hat{L}_k\,\hat {\cal H} \ ,
\end{eqnarray}
where ($i,j,k=x,y,z$); due to the presence of the quadratic term ($\hat{L}_k\,\hat {\cal H}$) in the rhs of the third commutation relation, they do not form a Lie algebra.

Instead of solving the Schr\"odinger equation
\[
  \hat {\cal H}\,\Psi\ =\ E\,\Psi\ ,
\]
we introduce the operator
\begin{equation}
\label{H}
  r\,(\hat {\cal H}\ -\ E)\ =\ \left(-\frac{r}{2} \De^{(3)}\, -\, E \,r\,\right)\ -\ \al \ \equiv (\hat {K} - \al)\ ,
\end{equation}
and look for its zero modes, here $\De^{(3)}$ is the three-dimensional Laplacian. Evidently, this is equivalent to the study of the spectrum of the operator $\hat K$,
\begin{equation}
\label{H-equation}
{\hat K} \,\Psi\ =\  \al \,\Psi\ ,\quad \hat K=\left(-\frac{r}{2} \De^{(3)}\, -\, E \,r\,\right)\ ,
\end{equation}
see e.g. \cite{TE:2023},
where $\al$ is the spectral parameter, while the energy $E$ - the original spectral parameter - is considered fixed. This procedure resembles the introduction of the so-called Sturm representation for the Hydrogen atom in quantum mechanics. Note that in spherical coordinates after separation of angular variables in $\hat K$ the resulting radial operator in variable $w={\sqrt r}$ corresponds to the radial Schr\"odinger operator of four-dimensional harmonic oscillator with frequency $\om^2=-E$.

By using the {\it coupling constant metamorphosis} formalism introduced in \cite{Hietarinta} and further established by Kalnins-Miller-Post \cite{Miller:2010}, it can be checked that for the operator ${\hat K}$, see (\ref{H-equation}), the angular momentum $\hat {\bf L}$ (\ref{L}) remains to commute with it but the Laplace-Runge-Lenz vector (\ref{A}) should be \textit{modified}
\begin{equation}
\label{B}
   \hat{\bf  B} \ = \ \frac{1}{2}(\,\hat{\bf  p} \times \hat{\bf  L}\,-\,\hat{\bf  L}
   \times \hat{\bf  p}\,) \ - \ \frac{\al}{r}\,{\bf r}\,\hat {\cal H}\ ,
\end{equation}
in order to maintain an integral of motion. Hence, these $\hat {\bf L}$ and $\hat {\bf B}$
are the integrals of motion for ${\hat K}$: $[\hat{\bf  L} ,\,{\hat K}]=[\hat{\bf  B} ,\, {\hat K}]=0$. It can be also verified that
\begin{equation}
\hat{\bf  B}^2 \ = \ {\hat K}^2 \ + \ 2 \,E\,(\,\hat{\bf L}^2 \ + \ 1\,)
\end{equation}
cf.(\ref{A2}), and
\begin{equation}
\hat{\bf  L}\,\cdot\,\hat{\bf B} \ = \ \hat{\bf  B}\,\cdot\,\hat{\bf L} \ = \ 0 \ .
\end{equation}
Thus, the operator ${\hat K}$ is {\it maximally superintegrable}. Let us note that the above integrals obey the commutation relations:
\begin{equation}
\label{ConmutRel}
[\hat{B}_i \,,\,\hat {L}_j] \ = \ i\,\epsilon_{ijk}\,\hat{B}_k \quad , \quad  [\hat{B}_i \,,\,\hat {B}_j] \ = \ -2\,i\,\epsilon_{ijk}\,\hat{L}_k\,E \ ,
\end{equation}
By adding the commutation relation $[\hat{L}_i \,,\,\hat {L}_j] \ = \ i\,\epsilon_{ijk}\,\hat{L}_k$, see (\ref{LA-comm-Rel}), one finds that for $E<0$ the operators $\hat {\bf L}, \hat {\bf B}$ span the Lie algebra $so(4)$.

\section{Two-dimensional problem}

\vspace{0.3cm}

Let us consider the expression
\begin{equation}
\label{gf}
\Gamma \ = \ \rho^{|m|}\,e^{-\beta\,r} \,z^p\,e^{i\,m\,\varphi} \ ,
\end{equation}
as a gauge factor, where $m=0,\pm1,\pm2\ldots$, is the magnetic quantum number, $p =0, 1$ defines parity in the $z$-direction, $\beta \equiv \sqrt{-2\,E}$, $\rho=\sqrt{x^2+y^2}$, $\tan(\varphi)=y/x$. We introduce the gauge rotated operator
\begin{equation}
{\hat h} \ \equiv \ \Gamma^{-1}\, {\hat K}\,\Gamma  \ ,
\end{equation}
with ${\hat K}$ defined in (\ref{H-equation}). Thus, the eigenfunctions of ${\hat K}$ are of the form $\Psi(r,\rho,\varphi)=\Gamma \,\psi(r,\,\rho)$. In this case, the variable $\varphi$ is easily separated out and the operator ${\hat h}$ has $\psi(r,\,\rho)$ as the eigenfunctions. After straightforward calculations we arrive at
\begin{eqnarray}
\label{ho}
{\hat h}(r,\,\rho)  & \  =   \ -\,\frac{1}{2}\,r\,\pa^2_r \ - \ \frac{1}{2}\,r\,\pa^2_\rho \ - \ \rho\,\pa^2_{r,\rho}\ - \ \frac{(1+2 |m|)\,r\,-\,2\,\beta\,\rho ^2}{2\, \rho }\, \pa_\rho \nonumber
\\ &
 -( 1+p+|m| \,-\,\beta\,r )\,\pa_r \ + \ \beta \,(1+p+|m|)  \ .
\end{eqnarray}
In the variables ($r,\,u \equiv \rho^2$), the operator ${\hat h}$ (\ref{ho}) takes the form of an algebraic operator with polynomial coefficients \cite{TE:2023},
\begin{eqnarray}
\label{h-algebraic}
\hspace{-0.0cm}{\hat h}_a(r,u) & \  =  \ -\,\frac{1}{2}\,r\,\pa^2_r \ - \ 2\,r\,u\,\pa^2_u \ - \ 2\,u\,\pa^2_{r,u}
\ - \ 2\,[(1+|m|)\,r\,-\,\beta\,u]\, \pa_u \non
\\ &
\hspace{-1.4cm} - \, ( 1+p+|m| \,-\,\beta\,r)\,\pa_r \ + \ \beta \,(1+p+|m|)\ ,
\end{eqnarray}
and the eigenvalue problem (\ref{H-equation}) becomes
\begin{equation}
\label{h-equation}
{\hat h}_a \,P\ =\  \al \,P\ ,
\end{equation}
here the domain is defined as $D=(r \geq 0, r \geq \rho \geq 0)$. {Its spectrum can be easily found
\[
   \al_n =\ \beta (n \ + \ 1+p+|m|)\ ,
\]
see \cite{TE:2023}.}

\vspace{0.2cm}


It is evident that the combination of integrals
\begin{equation}
   L_a \equiv {\hat L}_x^2\,+\,{\hat L}_y^2 \qquad {\rm and} \qquad
   B_a \equiv {\hat B}_x^2\,+\,{\hat B}_y^2 \ ,
\end{equation}
commutes with ${\hat L}_z$. It is also evident that the gauge-rotated two-dimensional operators
\begin{equation}
  {\hat l}_a \ \equiv \ \Gamma^{-1}\, ({\hat L}_x^2\,+\,{\hat L}_y^2)\,\Gamma
  \qquad {\rm and} \qquad
    {\hat b}_a \ \equiv \ \Gamma^{-1}\, ({\hat B}_x^2\,+\,{\hat B}_y^2)\,\Gamma
\end{equation}
commute with ${\hat h}_a$ (\ref{h-algebraic}):
\[
     [{\hat h}_a,\,{\hat b}_a]=0\quad ,\quad [{\hat h}_a,\,{\hat l}_a]=0\ .
\]
In explicit form,
\begin{equation}
\label{lop}
    {\hat l}_a \ = \ 2\,u\,(r^2\,-\,u)\,\pa^2_u \ - \ \big(\, u\,(1+2\,p) \ - \ 2\,(1+|m|)(r^2\,-\,u)\, \big)\,\pa_u \ ,
\end{equation}
is a second order differential operator, while
\begin{eqnarray}
\label{bop}
& \hspace{0.0cm}2\,{\hat b}_a  \, =  \, \frac{u}{4}\,\pa^4_r \, + \, 4 \,u^3\,\pa^4_u \, + \, 2\, u\,
    (2 r^2 + u)\,\pa^2_u \pa^2_r  \, + \, 2 \,r \,u\, \pa^3_r\,\pa_u \, + \,
    8\, r \,u^2\,\pa^3_u\,\pa_r  + \big((m+1)r-\beta \, u\big)\,\pa^3_r \nonumber
\\
&  + \ 8 u^2 (m+p+3-\beta  r)\,\pa^3_u \ +
   \ \bigg(2 u (m+p-3 \beta  r+3)+4 (m+1) r^2 \bigg)\,\pa^2_r\,\pa_u \nonumber
\\
& \hspace{-0.8cm}+ \ 4\, u\, \bigg(\,r (3 m+2 p-2 \beta  r+7)-\beta  u\bigg) \,\pa^2_u\,\pa_r \ +
  \ \bigg((m+1) (m+p+1)-3 \beta  (m+1) r+\beta ^2 u \bigg)\,\pa^2_r  \nonumber
\\
& \hspace{-0.5cm} + \ \bigg(4 \,r \,\bigg((m+1) (m+2 p+3)+\beta ^2 u \bigg)-4 \beta  u (m+p+3)
    - 8 \beta  (m+1) r^2 \bigg)\,\pa_r\,\pa_u \ + \nonumber
\\
& \hspace{-0.5cm} 4\,u\,\bigg(m^2+3 m (p-\beta  r+2)+p (7-2 \beta  r)+\beta  r (\beta  r-7)+7\bigg)\,\pa^2_u
    \, - \, 2 \beta  (m+1) (m+p+1-\beta  \,r)\,\pa_r \nonumber
\\
& \hspace{-0.7cm} \ + \ \bigg(\,4 (m+1) (p+1) (m+p+1) - 4 \beta  (m+1) r (m+2 p+3) + 2 \beta ^2 \left(2 (m+1) r^2+u\right)\,\bigg)\,\pa_u \ ,
\end{eqnarray}
is a fourth order differential operator. Hence, the operator ${\hat h}_a$ (\ref{h-algebraic}) is maximally superintegrable.

\subsection{$2D$ Schr\"odinger operator}

\vspace{0.2cm}

By using the gauge factor
\begin{equation}
\Gamma_a \ \equiv \  e^{\beta\,r}\,{(r^2-u)}^{-\frac{p}{2}}\,u^{-\frac{1+2|m|}{4}}\ ,
\end{equation}
the algebraic operator ${\hat h}_a$ (\ref{h-algebraic}) can be transformed into a two-dimensional Schrödinger operator
\begin{equation}
    {\cal H}^{(2)}_{\rm LB} \ \equiv \ \Gamma^{-1}_a \,\,{\hat h}_a\,\,\Gamma_a \ =
    \ -\De_{\rm LB}^{(2)} \ + \ V_{\rm eff}(r,u) \ ,
\end{equation}
where $\De_{\rm LB}^{(2)}$ is the two-dimensional Laplace-Beltrami operator,{
\[
\De_{\rm LB}^{(2)}=\frac{1}{\sqrt{g}}\,\pa_\mu {\sqrt{g}}\,g^{\mu\,\nu}\,\pa_\nu\ ,\quad \mu,\nu=r,u\ ,
\]
where $g={\rm{det}}[g_{\mu\,\nu} ]$ is the determinant of metric,
}with cometric
\begin{equation}
g^{\mu\,\nu} \ = \ \left|
 \begin{array}{cc}
 \frac{1}{2} \,r & \ u
     \\
u & \  2\,r\,u
 \end{array}
               \right| \ ,
\end{equation}
whereas the effective potential is given by
\begin{equation}
V_{\rm eff} \ = \ \frac{\left(4 |m|^2-1\right) \,r}{8\, u}\ + \ \frac{\beta ^2 }{2}\,r \ .
\end{equation}

The determinant of $g^{\mu\,\nu}$ reads ${\rm det}[g^{\mu\,\nu} ]=u\,(r^2-u)$, and the scalar curvature is $R=\frac{r \,(4 u-1)}{2 \,u\, \left(r^2-u\right)^2}$. Finally, we arrive at a new form of the Coulomb problem,
\begin{equation}
\label{H2}
  {\cal H}^{(2)}_{\rm LB} \ =
    \ -\De_{\rm LB}^{(2)} \ + \ \frac{\left(4 |m|^2-1\right) \,r}{8\, u}\ + \
    \frac{\beta ^2 }{2}\,r \ .
\end{equation}
This defines an exactly-solvable, maximally-superintegrable problem, see below.

\section{Polynomial algebra}

\vspace{0.2cm}

From (\ref{lop}) and (\ref{bop}), by taking the commutator we construct a fifth
order differential operator
\begin{eqnarray}
\label{C}
   & \hspace{-2.0cm}  {\hat c} \ \equiv \ [\,{\hat b}_a,\,{\hat l}_a\,]  \nonumber
   \\ & \hspace{-1.0cm}
   = \ 8\, u^3 \,\left(r^2-u\right)\,\pa_u^5 \ + \ 8 \,r\, u^2\, \left(r^2-u\right)\,\pa_u^4\,\pa_r \ + \ 2\, r\, u\, \left(u-r^2\right)\pa_u^2\,\pa_r^3 \ + \ \frac{1}{2} u\, \left(u-r^2\right)\pa_u\,\pa_r^4 \nonumber
\\ & \hspace{-1.0cm}
    + \ 2\, u^2\, \left(2 r^2 (5 m+2 p+17)-10 u (m+p)-4 \beta  r^3+4 \beta r u-37 u\right)\,\pa_u^4  \nonumber
\\ & \hspace{-1.0cm}
    + \ 8 \,r\, u\, \left(-2 u (m+p)+2 (m+2) r^2-5 u\right)\pa_u^3\,\pa_r \, + \, 3 u \left(-2 (p+1) r^2+2 \beta  r^3-2 \beta  r u+u\right)\pa_u^2\,\pa_r^2  \nonumber
\\ & \hspace{-1.0cm}
    - \ 2 \left(r^2-u\right) (m r+r-\beta  u)\pa_u\,\pa_r^3 \ + \ \frac{1}{8} \left(2 \,u \,(m+p)-2 (m+1) r^2+3\, u\right)\pa_r^4
     \ + \ {\cal O}(\pa^3)\ ,
\end{eqnarray}
which commutes with ${\hat h}_a$ (\ref{h-algebraic}).

Now, let us consider the algebra generated by the four elements $({\hat h}_a,\,{\hat l}_a,\,{\hat b}_a,\, {\hat c})$.
The relevant commutators are the following:
\begin{eqnarray}
& \hspace{-1.0cm} [\,{\hat c},\,{\hat l}_a\,]   \ =  \ 2\,{\hat l}_a\,{\hat h}_a^2
  \ -\ 8\,{\hat l}_a\,{\hat b}_a\
  \ - \ 4\,\beta^2\,{\hat l}_a^2\ - \  \,8 \,\beta  (3 m+2 p+7)\,{\hat l}_a\,{\hat h}_a  \nonumber
     \ - \ (m+1) (2 m+2 p-1)\,{\hat h}_a^2
\\ & \nonumber \hspace{-1.5cm}
     +  \ \,2\,\beta^2 \bigg(11 m^2+m (20 p+38)+(p+1)(9 p+26)\bigg)
     \,{\hat l}_a
\\ & \nonumber \hspace{-1.5cm}
   - \ 4\,{\hat c} \ + \ (2 m+2 p-1) (2 m+2 p+3)\,{\hat b}_a
\\ & \nonumber \hspace{-1.5cm}
  + \ \beta \,(2 m+2 p-1)\,  (2 m+2 p+3)\, (3 m+2 p+7) \,{\hat h}_a
\\ &  \hspace{-1.5cm}
 - \ \beta ^2\, (2 m+2 p-1) \,\bigg(\,m \,(m\, (5 m+14 p+26)+62 p+41)+66 p+20\bigg)\ ,
\end{eqnarray}
which is a third order polynomial in the generating elements, or equivalently, a sixth order differential operator, and
\begin{eqnarray}
 & \hspace{-1.0cm}[\,{\hat c},\,{\hat b}_a\,]   \ =\ - \ 4\, \beta ^2\,{\hat l}_a\,{\hat h}_a^2
 \ - \ 2\,{\hat b}_a\,{\hat h}_a^2 \ - \ 2 \,\beta\,  (3 m+2 p+7)\,{\hat h}_a^3  + \ 4\,{\hat b}_a^2\ + \ 8 \, \beta \,  (3 m+2 p+7)\,{\hat b}_a\,{\hat h}_a \nonumber
\\ & \nonumber \hspace{-0.0cm}
 + \ 8\, \beta ^2\,{\hat l}_a\,{\hat b}_a + \ 8 \,\beta ^3\, (3 m+2 p+7)\,{\hat l}_a\,{\hat h}_a \
 + \ 2 \,\beta ^2\, (m (21 m+30 p+94)+76 p+105)\,{\hat h}_a^2
\\ & \nonumber \hspace{-1.6cm}
- \ 2\, \beta ^2 \,(m (11 m+20 p+38)+44 p+26)\,{\hat b}_a \
\\ & \nonumber \hspace{-1.6cm}
  + \ 4\,\beta^2\,{\hat c}
 \ - \ 2 \,\beta ^3 \,\bigg(m (m (33 m+82 p+191)+4 (97 p+86))+448 p+182\bigg)\,{\hat h}_a
\\ & \nonumber \hspace{-1.6cm}
  - \ 4 \,\beta ^4 \,\bigg(5 m^2+2 m (4 p+9)+4 (p+1) (p+3)\bigg)\,{\hat l}_a
\\ & \nonumber \hspace{-0.0cm}
 + \ 2 \,\beta ^4\, (m+p+1) \bigg(15 m^3+m^2 (39 p+89)+3 m (p (74-11 p)+54)
\\ &  \hspace{-1.4cm} 
 +\ p\, (\,293-p (8 p+65)\,)+84\bigg)
 \ ,
\end{eqnarray}
which is a third order polynomial in the generating elements, or an eighth order differential operator. Hence, by definition we arrive at a polynomial algebra \cite{Bonatsos,Quesne} 
generated by
$({\hat h}_a,\,{\hat l}_a,\,{\hat b}_a,\, {\hat c})$. In our case, it is a {\it cubic} polynomial algebra.

\section{Lie-algebraic structure}

\vspace{0.2cm}

It can be immediately checked that the operator ${\hat h}_a$ (\ref{h-algebraic}) has infinitely-many finite-dimensional invariant subspaces
\begin{equation}
\label{Pn}
   {\mathcal P}_n\ =\ \langle \, r^p u^q\ |\ 0 \leq p+2q \leq n\, \rangle \ ,
\end{equation}
\[
   {\hat h}_a:\,{\mathcal P}_n \ \rar {\mathcal P}_n \ ,\qquad n=0,1,2,\ldots \ ,
\]
which form the infinite flag.
For fixed $n$ the invariant subspace ${\mathcal P}_n$ coincides with the finite-dimensional representation space of the algebra $g^{(2)}$: the infinite-dimensional, eleven-generated algebra of differential operators. This algebra was introduced in \cite{Turbiner:1998} (see for a discussion \cite{Turbiner:2013t}) in relation to the $G_2$-integrable system of the Hamiltonian reduction. It is spanned by the Euler-Cartan generator
\begin{eqnarray}
\label{J0-g2}
{\tilde {\cal J}}_0(n)\ =\ r\,\pa_r \  +\ 2 \,u\,\pa_u \ - \ n\ ,
\end{eqnarray}
and the generators\[
 \hspace{-2.3cm} {\cal J}^1\  =\  \pa_r\ ,\
 {\cal J}^2_n\  =\ r \,\pa_r\ -\ \frac{n}{3} \ ,\
 {\cal J}^3_n\  =\ 2\, u\,\pa_u\ -\ \frac{n}{3}\ ,
\]
\begin{eqnarray}
\label{gl2r}
   &   \hspace{-2.3cm}  {\cal J}^4_n\  =\ r^2 \pa_r \  +\ 2 \,r\, u\, \pa_u \ - \ n r\ = \ r {\tilde {\cal J}}_0(n)\ ,
\\ &   \hspace{-2.3cm}  {\cal R}_{0}\  = \ \pa_u\ ,\ {\cal R}_{1}\  = \ r\,\pa_u\ ,\ {\cal R}_{2}\  = \ r^{2}\,\pa_u\ , \nonumber
\end{eqnarray}
\begin{eqnarray}
\label{gl2t}
\hspace{0.0cm} {\cal T}_0\ =\ u\pa_{r}^2 \ ,\ {\cal T}_1\ =\ u\pa_{r} {\tilde {\cal J}}_0{(n)}\ ,\
 {\cal T}_2\ =\
  u{\tilde {\cal J}}_0{(n)}\ ({\tilde {\cal J}}_0{(n)} + 1) = \ u {\tilde {\cal J}}_0{(n)}\ {\tilde {\cal J}}_0{(n-1)}\ ,
\end{eqnarray}
where $n$ is a parameter, which has a meaning of the mark of the representation. If $n$ is a non-negative integer, the space ${\mathcal P}_n$ is the common invariant subspace for the generators (\ref{J0-g2}), (\ref{gl2r}), (\ref{gl2t}), where the algebra acts irreducibly.

The differential operator ${\hat h}_a(r,u)$ can be written in the $g^{(2)}$-Lie-algebraic form, see \cite{TE:2023},
\begin{equation}
\label{h-Lie-algebraic}
 {\hat h}^{(g^{(2)})}_{a}\ =\ -\,\frac{1}{2}\,{\cal J}^2_0 {\cal J}^1\ - \ {\cal J}^3_0\,
 {\cal R}_{1} \ - \ {\cal J}^3_0\,{\cal J}^1\ +\ \beta {\tilde {\cal J}}_0(0)\ -
\end{equation}
\[
   \left( 1+p+|m|\right)\,{\cal J}^1\ - \ 2\,\left(1+|m|\right)\,{\cal R}_{1}
   \ + \ \beta \,(1+p+|m|)\ ,
\]
in terms of the generators (\ref{J0-g2}), (\ref{gl2r}) (at $n=0$) only. {In a similar way the integrals ${\hat l}_a$  (\ref{lop}),
\begin{equation}
\label{la-Lie-algebraic}
    {\hat l}_a \ = \ {\cal J}^3_0\,{\cal R}_{2}\,-\,\frac{1}{2}\,{\cal J}^3_0\,{\cal J}^3_0\,
    \ - \,\frac{1}{2}\,(1+2\,p+2|m|)\,{\cal J}^3_0\ +\ 2\,(1+|m|){\cal R}_{2} \ ,
\end{equation}
as well as ${\hat b}_a$ (\ref{bop}) and their commutator}
${\hat c} \ = \ [\,{\hat b}_a,\,{\hat l}_a\,]$ (\ref{C}) admit a representation in terms of the
generators (\ref{J0-g2}), (\ref{gl2r}).

\noindent
{\it Remark.}
Note that the generators (\ref{J0-g2}), (\ref{gl2r}) span the subalgebra $gl(2) \ltimes {\it R}^3 \in g^{(2)}$,
discovered by Sophus Lie around 1880 \cite{Lie:1880} as the algebra of vector fields,  see \cite{gko:1992} (Cases 24, 28 at $r=2$), which was extended to first order differential operators in \cite{gko:1994}. In general, this algebra (\ref{J0-g2}), (\ref{gl2r}) acts on functions of two variables $(r,u)$. However, in the action on functions of the single variable $r$, this algebra acts as a $sl(2)$-algebra,
\begin{equation}
\label{sl2}
  J^+_n = r^2\pa_r - n \,r\ ,\ J^0_n = 2\,r\,\pa_r - n\ ,\ J^- = \pa_r \ .
\end{equation}

\section*{Conclusions}

In this work we showed that the two-body Coulomb problem in the Sturm representation
leads to a new two-dimensional, exactly-solvable, superintegrable quantum system
in curved space with variable curvature, which has $g^{(2)}$ hidden algebra and possesses
a cubic polynomial algebra of integrals. Two integrals are built
from the two components of angular momentum and from the modified Laplace-Runge-Lenz vector, respectively.
The Hamiltonian and these integrals can be written as algebraic operators with polynomial
coefficients, they admit a $g^{(2)}$-Lie-algebraic form: they can be represented as a non-linear combinations of the $g^{(2)}$-algebra generators.
The cubic polynomial algebra of integrals is an infinite-dimensional subalgebra of the universal enveloping algebra $U_{g^{(2)}}$. This confirms a suspicion, known in folklore (P.~Winternitz,
W.~Miller Jr, A. V.~Turbiner), that if a maximally superintegrable system has a hidden algebra, the polynomial algebra of integrals belongs to the universal enveloping algebra of the hidden algebra, usually, this system is exactly-solvable (the so-called Montreal conjecture), see \cite{TTW:2001}.

\section*{Acknowledgments}

\noindent
This work is partially supported by CONACyT grant A1-S-17364 and DGAPA grant IN113022 (Mexico). A.M.E.R. thanks the audience who raised the questions about integrals and possible polynomial algebra during his presentation at the XII International Symposium on Quantum Theory and Symmetries (QTS12) in Prague.

\end{document}